\documentclass[aps,preprint,preprintnumbers, amsmath, amssymb, prb]{revtex4}
\usepackage{times}
\usepackage{graphicx}
\usepackage{psfrag}
\usepackage{ae}
\usepackage{amsmath,amssymb}

\begin{document}
\date{\today}

\author{J. R. Bordin} 
\email{bordin@if.ufrgs.br}
\affiliation{Programa de P\'os-Gradua\c c\~ao em F\'{\i}sica, Instituto 
de F\'{\i}sica, Universidade Federal
do Rio Grande do Sul\\ Caixa Postal 15051, CEP 91501-970, 
Porto Alegre, RS, Brazil}

\author{A. B. de Oliveira}
\email{oliveira@iceb.ufop.br}
\affiliation{Departamento de F\'{\i}sica, Universidade Federal de Ouro Preto, Ouro 
Preto, MG, 35400-000, Brazil }

\author{A. Diehl}
\email{diehl@ufpel.edu.br}
\affiliation{Departamento de F\'{\i}sica, Instituto de F\'{\i}sica e Matem\'atica,
  Universidade Federal de Pelotas, Caixa Postal 354, CEP 96010-900, Pelotas, RS, Brazil}

\author{Marcia  C. Barbosa} 
\email{marcia.barbosa@ufrgs.br}
\affiliation{Instituto de F\'{\i}sica, Universidade Federal
do Rio Grande do Sul\\ Caixa Postal 15051, CEP 91501-970, 
Porto Alegre, RS, Brazil}

\title{Diffusion Enhancement in Core-softened fluid confined in nanotubes}

\begin{abstract}
We study the effect of confinement in the dynamical behavior of a
core-softened fluid.  The fluid is modeled as a two length
scales potential. This potential in the bulk  reproduces the 
anomalous behavior observed in the density
and in the diffusion of  liquid water.
A series of $NpT$ Molecular Dynamics simulations for this two
length scales fluid  confined in a nanotube were performed. We obtain that the diffusion
coefficient increases with the increase of the  nanotube radius for wide  channels
as expected for normal fluids. However, 
for narrow channels, the confinement shows an enhancement in the diffusion coefficient
 when
the nanotube radius decreases.
This behavior, observed for water, is explained in the 
framework of the two length scales potential.

\end{abstract}


\maketitle
\section{Introduction}

The dynamic behavior of fluids in the bulk is characterized by
transport properties such as the diffusion coefficient. In
simple liquids they are governed by the molecular interactions which 
can be effectively treated as pair potentials.
In complex fluids, such  as water, the dynamics of the systems
is governed by network-forming interactions. The potentials involved
are in general highly anisotropic and the transport properties
tend to exhibit unusual behaviors.

Under confinement even normal liquids have an unusual behavior, very different from the
physical properties observed in bulk. The competition 
between surface effects and the confinement can
induce a dramatic change in the transport properties of the 
fluid inside the channel~\cite{Mao00, Liu10,Striolo06,Pikunic03, MicroNano05}.

Bulk water is anomalous in many of its characteristics. 
The maximum in water's density  is a  well-known anomaly  but there are many others. 
The self-diffusion coefficient at fixed temperature for a normal liquid decreases
under compression, while in liquid water it increases with the increase of 
pressure. In bulk water this is due to the 
hydrogen bonds that are created and destroyed 
making particles to move from one neighbor to another
neighbor.

Notwithstanding its
molecular simplicity water   is quite hard to be modeled.
The reason behind this difficulty is the presence
of the hydrogen bonds, a non symmetric charge distribution
and polarizability of the molecule that are density and temperature
dependents.
Consequently, there are more than twenty-five (bulk) water models 
for computational simulation -- empirical potentials -- 
in which each of them
give a different dipole moment, dielectric and self-diffusion constants, average 
configurational energy, density maximum, 
and expansion coefficient. More specifically, the maximum of density is
found experimentally to be at $T=4$ $^{\rm o}$C (for pressure $P=1$ atm) while 
such models give values ranging 
from -45 $^{\rm o}$C (SPC model) up to 25 $^{\rm o}$C (POL5/TZ model). The 
TIP5P water 
model was built to 
match the  4 $^{\rm o}$C experimental result, but it fails in many 
other aspects.\cite{watermodels}
Despite these limitations, these models have
been used to understand the 
transport properties and phase transitions of confined 
water~\cite{Gallo03,Brovchenko03,Giovambattista09,Han10,B805361H,
Franzese11,Gallo12,Strekalova12,Melillo11}.
The results  give  a qualitative
comparison with experiments without providing a complete  
understanding of the origin of the  anomalies.\cite{Mallamace12}
The majority of the molecular water models are conceived focusing on
accurately describe the hydrogen bonds and charge 
distributions since many of the water 
uncommon properties are believed to come 
from its highly directional interactions. Examples are
solvation and properties which depend on polarization.
On the other hand, the literature have many examples in which systems with absence of anisotropic interactions still may present some of the water features~\cite{Oliveira06a, Oliveira06b,Silva10,Oliveira10,
Barraz09,Oliveira08,Oliveira08b,Oliveira09, Er06, Mi06a, Kr08,xu:054505}.
Some of its anomalous behavior may 
come from purely volumetric effects, which particularly is our focus in this work.

For confined systems, where water molecules 
interact in nanoscale distances, first principles simulations would be
the appropriated tool for numerical comparison with
experimental data. This procedure, however, has limitations. Even for confined water 
systems, in which the sizes involved are much smaller than that ones found in bulk cases, 
thousands of atoms are necessary for attacking typical problems along with millions 
of simulation steps.
In this sense, turns out that in the majority of cases \emph{ab initio} techniques 
become impracticable for dealing with such computational 
demanding systems.

Given the limitations of the full water models and 
the computational costs of the \emph{ab initio} simulations, classical effective empirical potentials became the 
simplest framework to understand the physics behind the 
anomalies of bulk water. From the desire of constructing 
a simple two-body potential capable of 
describing the anomalous behavior of bulk water,
a number of models have been developed~\cite{Oliveira06a, Oliveira06b,Silva10,Oliveira10,
Barraz09,Oliveira08,Oliveira08b,Oliveira09, Er06, Mi06a, Kr08,xu:054505}.
Despite their simplicity, such models had successfully reproduced 
the thermodynamic, dynamic, and structural anomalous behavior
present in bulk liquid water. They also predict the existence of a second critical 
point hypothesized by Poole and collaborators~\cite{Po92}. This suggests
that some of the unusual properties observed in water can be quite universal and 
possibly present in other systems.

In the case of confined water a number of attempts have been
made to understand its thermodynamic and dynamic properties.
For the confinement media, nanotubes have been widely used for mimicking
water confined into live organisms and as building blocks
for technological applications, as desalination of water~\cite{Elimelech11}.
Also,  nanotubes can be used for drug delivery since they resemble biological ion channels~\cite{Hilder11}.
In addition, confinement in nanopores and nanotubes 
have been used also to avoid spontaneous water crystallization below
the melting point in an attempt to observe its hypothetical
second critical point~\cite{Liu05,Mallamace10,Chen06, Lombardo08, Stanley11}.
Simulations employing some of the discussed molecular models
for water, namely SPC/E, TIP4P-EW and ST2, confined
in nanoscale channels exhibit   two complementary effects: the 
melting temperature of the fluid
at the center of the channel decreases and water crystallizes at the 
channel
surface~\cite{Gelb99,Mashl03,Gordillo00,Kyakuno11, Alexiadis08}. 
In addition to these thermodynamic properties, the mobility properties
of confined water also exhibit an unusual behavior. Experiments
show an enhancement of the mobility orders of 
magnitude higher than what is predicted by the flow theories~\cite{Holt06,Majumder05}.
Simulations~\cite{Thomas08,Thomas09,Qin11} show
an increase in the enhancement  rate below a certain threshold radius.
Similarly the
self-diffusion coefficient, $D$,
obtained through molecular dynamic
simulations for atomistic models,
below a certain radius increases with
with decreasing radius~\cite{Mashl03, Ye11, Farimani11,Zheng12}.

Besides the thermodynamic and dynamic unusual properties of
confined water, the structure also presents
an interesting behavior.
The water structure inside larger nanotubes exhibits a 
layered structure, while in narrow nanotubes
a single file is observed. The layered water molecules can be found in
a spiral-like chain,~\cite{Liu05} in a hexagonal structure for (6,6) 
carbon nanotubes (CNT),~\cite{Mashl03}
a octagonal water-shell structure for a (9,9) CNT or 
a octagonal water-shell structure
 with a central water chain for a (10,10) CNT,~\cite{Koles06,Kolesnikov04} and others 
different structures~\cite{Alexiadis08}.

The presence  of layering effects
is also controversial.  Molecular Dynamics (MD) simulations results by 
Wang {\it et al.}~\cite{Wang04} do not show any obvious
ordered water structure for (9,9) or (10,10) CNTs, unlike 
the works of Koles {\it et al}~\cite{Koles06, Kolesnikov04}.
This difference in findings can  
be caused by the influence of the
water model used and corresponding Lennard-Jones parameters~\cite{Wang04}.
The molecular structure of water confined in nanotubes and the diffusion 
can be very different depending on the chosen water 
model  to perform the simulations~\cite{Alexiadis08_2}.
This difference arises as a consequence
of the fact that the water models used in 
classical all-atoms MD simulations,
like SPC/E, TIP3P, TIP4P, etc, are parametrized for bulk simulations, 
and for reproduce only few aspects of real water. 
So this models may show errors to describe the correct water behavior 
under strongly confinement, like inside nanotubes.

The behavior of the diffusion coefficient with
the nanotube radius is still under debate.
Some simulations show  a monotonical decrease of $D$ 
with decreasing nanotube radius~\cite{Liu08,Nanok09,Liu05_b} while other 
simulations indicate
the presence of a minimum~\cite{Mashl03, Ye11, Farimani11,Zheng12}.  
It has been suggested that the length of the 
nanotube and the length of the simulation
would be responsible for the different results~\cite{Striolo06,Mukherjee07,Su11} and 
that friction should play also a relevant role~\cite{Falk10}.

Our core-softened model introduced
to study bulk system  does not
have any directionality and therefore it is not water. However,
it does exhibit  the density, the diffusion and
the response functions anomalies observed in water. 
This  suggests that some of the anomalous properties that 
are attributed  directionality of
 water can be found in spherical symmetry systems. 
Likewise, here we propose that also the
minimum in the diffusion coefficient as 
the nanotube radius is decreased observed 
for water can be also found in spherical symmetric systems.

 In order to check our hypothesis, we model a water-like fluid
into a nanotube using a core-softened  potential. We test if 
this model is capable to capture 
the increase in the diffusion coefficient when the
channel radius decreases. Next, we verify if 
the layering and the structure formed inside the channel, 
observed in some classical models for confined liquid water, has
an universal feature or if it is just a consequence of the 
specific confining surface and model details.

The paper is organized as follows. The nanotube, water-like fluid 
model and the 
simulational details are presented in Sec.~\ref{Model}. Our results 
are discussed in Sec.~\ref{Results},
and the conclusions and summary are presented in Sec.~\ref{Conclu}.

\section{The Model and the Simulation details}
\label{Model}

\subsection{The Model}

\begin{figure}[ht]
\begin{center}
\includegraphics[width=8cm]{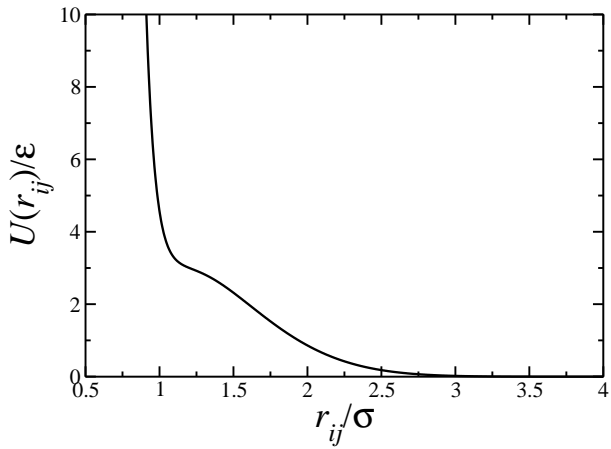}
\end{center}
\caption{Interaction potential between water-like particles.}
\label{fig:potential}
\end{figure}

The water-like fluid is modeled as point particles with effective diameter 
$\sigma$ and mass $m$, interacting through the three 
dimensional core-softened potential~\cite{Oliveira06a, Oliveira06b}
\begin{equation}
\frac{U(r_{ij})}{\epsilon} = 4\left[ \left(\frac{\sigma}{r_{ij}}\right)^{12} - 
\left(\frac{\sigma}{r_{ij}}\right)^6 \right] + 
u_0 {\rm{exp}}\left[-\frac{1}{c^2}\left(\frac{r-r_0}{\sigma}\right)^2\right]\;.
\label{AlanEq}
\end{equation}
The first term in Eq. (\ref{fig:potential}) is the standard Lennard-Jones (LJ) 
12-6 potential~\cite{AllenTild} and the second term is a Gaussian 
centered in $r_0/\sigma$,
with depth $u_0\epsilon$ and width $c\sigma$. For $u_0 = 5.0$, $c = 1.0$ 
and $r_0/\sigma = 0.7$ this equation 
represents a two length scale potential, with one scale 
at  $r_{ij}\approx 1.2 \sigma$, when the 
force has a local minimum, and the other scale at  $r_{ij} \approx 2 \sigma$, where
the fraction of imaginary modes has a local minimum~\cite{Oliveira10}.
  de Oliveira \emph{et al.}~\cite{Oliveira06a, Oliveira06b} obtained the 
pressure-temperature phase 
diagram of this system and showed that it exhibits thermodynamic, dynamic and 
structural anomalies  similar to the anomalies present in water~\cite{Kell67,Angell76}.

\begin{figure}[ht]
\begin{center}
\includegraphics[width=12cm]{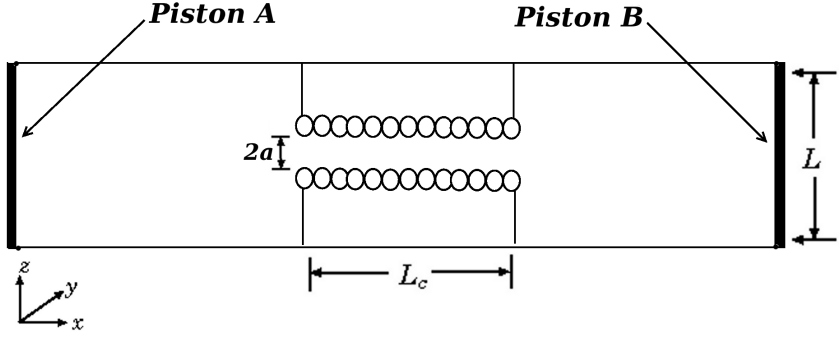}
\end{center}
\caption{Schematic depiction of the simulation box with the nanotube, reservoirs and fluctuating walls. 
The cylindrical channel in the center has radius $a$ and length $L_c$. The reservoirs  have height $L$.}
\label{fig:simulation-box}
\end{figure}

Here we study the dynamic behavior of this water-like model 
confined in a nanotube connected to two reservoirs. The 
nanotube-reservoir setup is 
illustrated in Fig.~\ref{fig:simulation-box}. The simulation box is a  parallelepiped with dimensions $L_x\times L_y\times L_z$. 

Two fluctuating walls, A in left and B in right, are placed in the limits of the $x$-direction of the simulation box.
The walls  are allowed to move in order to maintain the pressure constant in the reservoirs. The
 sizes  $L_y$ and  of $L_z$ depend on the effective nanotube 
radius, $a$, and they are  defined by
$L_y = L_z = L = 2a + 6\sigma$. The initial size $L_x$ is
given by  $L_x = 6L_c$, where $L_c$ is the tube length. 
The nanotube structure was constructed as a wrapped
hexagonal lattice sheet of point particles whose diameter is $\sigma_{\rm NT} = \sigma$. The nanotube interacts with the water-like particles
 through the Weeks-Chandler-Andersen (WCA) potential~\cite{AllenTild} given by 
\begin{equation}
\label{LJCS}
U_{ij}^{\rm{WCA}}(r) = \left\{ \begin{array}{ll}
U_{{\rm {LJ}}}(r) - U_{{\rm{LJ}}}(r_c)\;, \qquad r \le r_c\;, \\
0\;, \qquad \qquad \qquad \qquad \quad r  > r_c\;,
\end{array} \right.
\end{equation}
where $U_{\rm LJ}(r)$ is the standard LJ potential. The cutoff distance 
for this interaction is $r_c = 2^{1/6}\sigma_{ij}$, 
where $\sigma_{ij} = (\sigma_i + \sigma_j)/2$ is the center-to-center 
distance between the fluid particle $i$
and the nanotube particle $j$.

\subsection{The simulation details}

The properties of the system were evaluated with simulations at constant number of particles, pressure and temperature 
($NpT$ ensemble). The Andersen thermostat~\cite{Andersen80}  with collision frequency $\nu \delta t = 0.01$ 
was used to maintain the temperature fixed. The pressure in both reservoirs was fixed using the Lupowski and van Smol method of 
fluctuating confining walls~\cite{LupSmol90, Thompson99}.
These fluctuating walls act like pistons  in the system where 
a constant force controls the pressure in the $x$-direction.
This lead us to rewrite the resulting force in
a water-like particle as
\begin{equation}
 \vec F_R = -\vec\nabla U_{ij} + \vec F_{iwA}(\vec r_{iA}) + \vec F_{iwB}(\vec r_{iB})\;,
\end{equation}
where  $\vec F_{iwj}$ indicates the interaction between the 
particle $i$ and the piston $j$. These forces were calculated from a WCA potential similar to Eq.~(\ref{LJCS}),
however considering the $x$-projection of the distance between one particle in the bulk and the piston position. 

The equation of motion for the pistons are

\begin{equation}
 m_w\vec a_A = pS_w\vec n_A - \sum_{i=1}^N \vec F_{iwA}(\vec r_{iA})
\end{equation}
and
\begin{equation}
 m_w\vec a_B = pS_w\vec n_B - \sum_{i=1}^N \vec F_{iwB}(\vec r_{iB})\;,
\end{equation}
where $m_w$ is the piston mass, $p$ the desired pressure in the 
system, $S_w$ is the piston area and $\vec n_A$ is a unitary vector 
in positive $x$-direction, while $\vec n_B$ is a negative unitary vector. Both 
pistons (A and B) have
mass $m_w=m=1$, width $\sigma_w^x = \sigma$ and area equal to $S_w = L^2$.
The Andersen thermostat is also applied to the pistons to 
ensure the temperature control.
The values of pressure and temperature were chosen avoiding the 
density anomaly  and the solid state regions~\cite{Oliveira06a, Oliveira06b}.

For simplicity, we assume that the nanotube 
atoms are fixed (i.e., not time integrated) during the 
simulation. The reduced quantities are defined as usual,  
\begin{equation}
\label{red1}
a^*\equiv \frac{a}{\sigma}\;,\quad \rho^{*}\equiv \rho \sigma^{3}\;, \quad 
t^* \equiv t\left(\frac{\epsilon}{m\sigma^2}\right)^{1/2}
\quad \mbox{and}\quad T^{*}\equiv \frac{k_{B}T}{\epsilon}\;,
\end{equation}
for the channel radius, density of particles, time and temperature, respectively, and
\begin{equation}
\label{rad2}
p^*\equiv \frac{p \sigma^{3}}{\epsilon} \quad \mbox{and}\quad 
D^*\equiv \frac{D(m/\epsilon)^{1/2}}{\sigma}
\end{equation}
for the pressure and diffusion coefficient, respectively. 
Periodic boundary conditions were applied in the $y$ and $z$ 
directions. The equations of motion for the particles of the fluid were integrated using the velocity Verlet algorithm,
with a time step $\delta t^* = 0.005$.
The fluid-fluid interaction, Eq.~(\ref{AlanEq}), has a cutoff radius $r^*_{\rm cut} = 3.5$.
The nanotube radius was varied from $a^* = 1.25$ to $a^*=10.0$, and the 
number of fluid particles in the simulations varies from 500 to 3500.
The number of particles were chosen considering that the nanotube would be filled with the fluid and that we would have in the reservoirs the same properties evaluated in previous $NVT$ simulation for the non-confined case~\cite{Oliveira06a, Oliveira06b}.
For all values of radius the nanotube length was defined as 
$L_c^* = 20$.

Five independent runs were performed to evaluate the properties of the
fluid inside the nanotube. For each simulation run half of fluid particles was initially placed into each reservoir.
We performed $5\times10^5$ steps to equilibrate the system followed 
by $5\times10^6$ steps for the results production stage. The equilibration time
was taken in order to ensure that the  nanotube became filled with 
water-like particles as well as the pistons reached the equilibrium position for a given pressure.

For calculating the axial diffusion coefficient, $D_x$, we computed 
 the axial mean square displacement (MSD) namely
\begin{equation}
\label{r2}
\langle [x(t) - x(t_0)]^2 \rangle =\langle \Delta x(t)^2 \rangle=2Dt^\alpha\;,
\end{equation}
where $x(t_0)$ and  $x(t)$ denote the axial
coordinate of the confined water-like molecule 
at a time $t_0$ and a later time $t$, respectively.
The diffusion coefficient $D_x$ is then obtained from
\begin{equation}
 D_x = \lim_{t \to \infty} \frac{\langle \Delta x(t)^2 \rangle}{2t^\alpha}\/.
\end{equation}
Depending on the scaling law
between $\Delta x^2$ and $t$ in 
the limit $t \rightarrow \infty$, different diffusion mechanisms can be identified:
$\alpha=0.5$ identifies a single file regime~\cite{Farimani11}, 
$\alpha=1.0$ stands for a Fickian diffusion whereas $\alpha=2.0$ refers to a 
ballistic diffusion~\cite{Alexiadis08, Mukherjee07, Striolo06, 
Zheng12, Farimani11}.
 
\section{Results and Discussion}
\label{Results}
         
\begin{figure}[ht]
\begin{center}
\includegraphics[width=8cm]{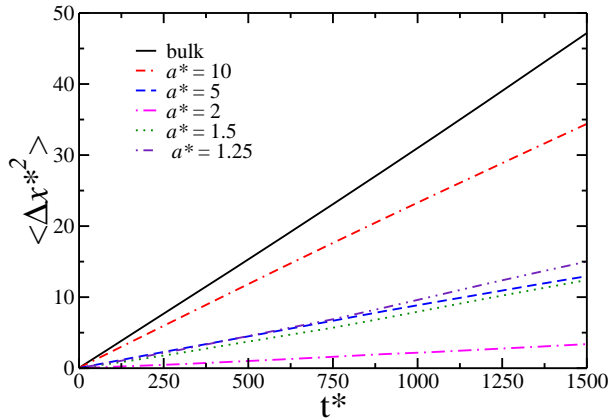}
\end{center}
\caption{Axial mean square displacement versus time for channel radius $a^*=1.25$, 1.5, 2.0, 5.0, and 10.0.}
\label{fig:x2.vs.t}
\end{figure}

First, we checked  which is  the diffusive regime of our system 
for different channel radius. 
Fig.~\ref{fig:x2.vs.t} illustrates the axial mean square 
displacement versus time for channel radius 
$a^*=1.25$, 1.5,  2.0, 5.0, and 10.0 at $T^*=0.25$ and $p^*=0.7$. 
For simple LJ confined fluids for very 
narrow channels a  single-file diffusion  
regime is found ($\alpha=0.5$)~\cite{Striolo06,Pikunic03}.
Instead, in our model Fickian diffusion was observed for
all channel radius,  ($\alpha=1.0$).  
This result is in agreement with the diffusion coefficient at
the center of nanotubes
 observed for  water 
models SPC/EP~\cite{Striolo06,Mashl03} , SPC~\cite{Liu08}, TIP3P~\cite{Alexiadis08_2} and TIP4P~\cite{Zheng12}.
In the case of water the Fickian diffusion has 
been interpreted as a combination of highly coordinated 
like in a ballistic mode and a monodimensional configuration
like in a single-line diffusion~\cite{Mukherjee07,Alexiadis08}. 
 The presence of
a highly coordinated structure was observed in water in very narrow
nanotubes~\cite{Farimani11} and between plates~\cite{Sa12}, confirming
the interpretation of Mukherjee et al.~\cite{Mukherjee07}.
In our case, the particles also combine a strongly correlated motion 
with a single-line diffusion. As we are going to see in 
detail below, the molecules are arranged in shells. The motion
inside each cell is correlated and slow while the motion between
shells is faster. The compromise
between these two mechanisms lead to a Fickian diffusion.

It is important to stress that the major difference between 
a simple LJ fluid and our core-softened model is the presence of
two-scales in the potential (absent in the LJ fluid). This enforces
our conjecture that two scales in the interatomic potential plays an 
important role in the appearance of water-like features.

\begin{figure}[t]
\begin{center}
\includegraphics[width=8cm]{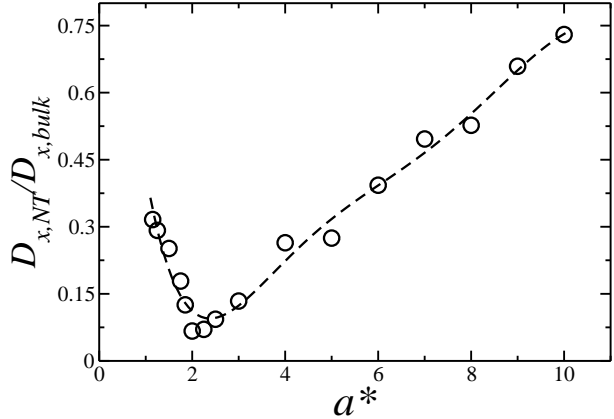}
\end{center}
\caption{Diffusion coefficient inside nanotube, $D_{x,NT}$, in units 
of non-confined diffusion, $D_{x,bulk}$, for different nanotube radius. 
The error bars are smaller than the data point. The dotted line is a guide to the eye.}
\label{fig:D.vs.a}
\end{figure}

Next, we tested  if the diffusion through the channel
obeys the mean-field-like Knudsen equation, i.e.,
if the diffusion coefficient is proportional to the channel radius.
Fig.~\ref{fig:D.vs.a} illustrates the diffusion coefficient, $D$, 
versus channel radius, $a^*$, for fixed $T^*=0.25$ and $p^*= 0.7$. 
We see from this figure that a critical channel radius $a^*_c$ exists 
where the derivative of the $D(a)$ curve is zero. 
For $a^*> a^*_c = 2.0$ the diffusion
coefficient presents the expected behavior of increasing with $a^*$. For large channel radius the growth is linear  
as predicted by the Knudsen equation. For
$a^*<2.0$, on the other hand, we observed that $D$ decreases with increasing $a^*$, which 
can not be explained by the Knudsen mean-field
approach. At $a^*= a^*_c = 2.0$ particles are virtually immobilized, i.e., $D\approx 0$.

Studies for SPC/E~\cite{Mashl03,Farimani11,Nanok09},  TIP4P-EW~\cite{Ye11,Zheng12} 
and SPC~\cite{Liu05_b, Liu08} show that the diffusion coefficient
increases with the channel radius, $a$. 
Simulations for  SPC/E~\cite{Mashl03,Farimani11} and  
TIP4P-EW~\cite{Ye11,Zheng12} also show the decrease of the diffusion 
coefficient with the increase of the channel radius 
for $a<a_c$. This anomalous region is captured by our 
model and it is not observed
in our simulations for LJ confined 
fluids (details not shown here for simplicity). 

For SPC/E and  TIP4P-EW potentials used for confined water, 
the number of neighbors and
the number of hydrogen atoms  differ from those numbers in the bulk
phase. In these models the different slope in the $D(a)$ function,
i.e., positive for $a>a_c$ and negative for $a<a_c$, are attributed to a 
competition between two effects: the confinement and 
the nanoscale surface. For the  $a>a_c$ case,  $D$ decreases for 
decreasing $a$ because 
of the confinement. 
This is not hard to understand since decreasing 
$a$ allows less space for particles to move~\cite{Mashl03, Ye11, Farimani11,Zheng12}. 
Increasing confinement leads to surface effects becoming more 
important. In the water case hydrogen bonds 
from the surface are depleted and molecules become more mobile. This would 
explain why $D$ increases for 
decreasing $a$ below $a_c$~\cite{Mashl03,Ye11,Farimani11,Zheng12}.
The behavior of $D$ in water, therefore, can be 
explained by minimizing the free energy. Water molecules
have a gain in rotational entropy when they are come inside
a narrow carbon channel~\cite{Ku11,Pa11}. This gain gain in entropy compensates
the lost in elthalpy due to the reduction of number of hydrogen bonds~\cite{Farimani11,Franzese11}. 

Interestingly though our system does not have
hydrogen bonds,  
thus it is not subject to the competition between hydrogen bonds
depletion and diminishing space available for particles to move.
Therefore, what would be
the mechanism behind  our non-monotonic curve $D(a)$ shown
in Fig.~\ref{fig:D.vs.a}?
                                 
\begin{figure}[ht]
\begin{center}
\includegraphics[width=5cm]{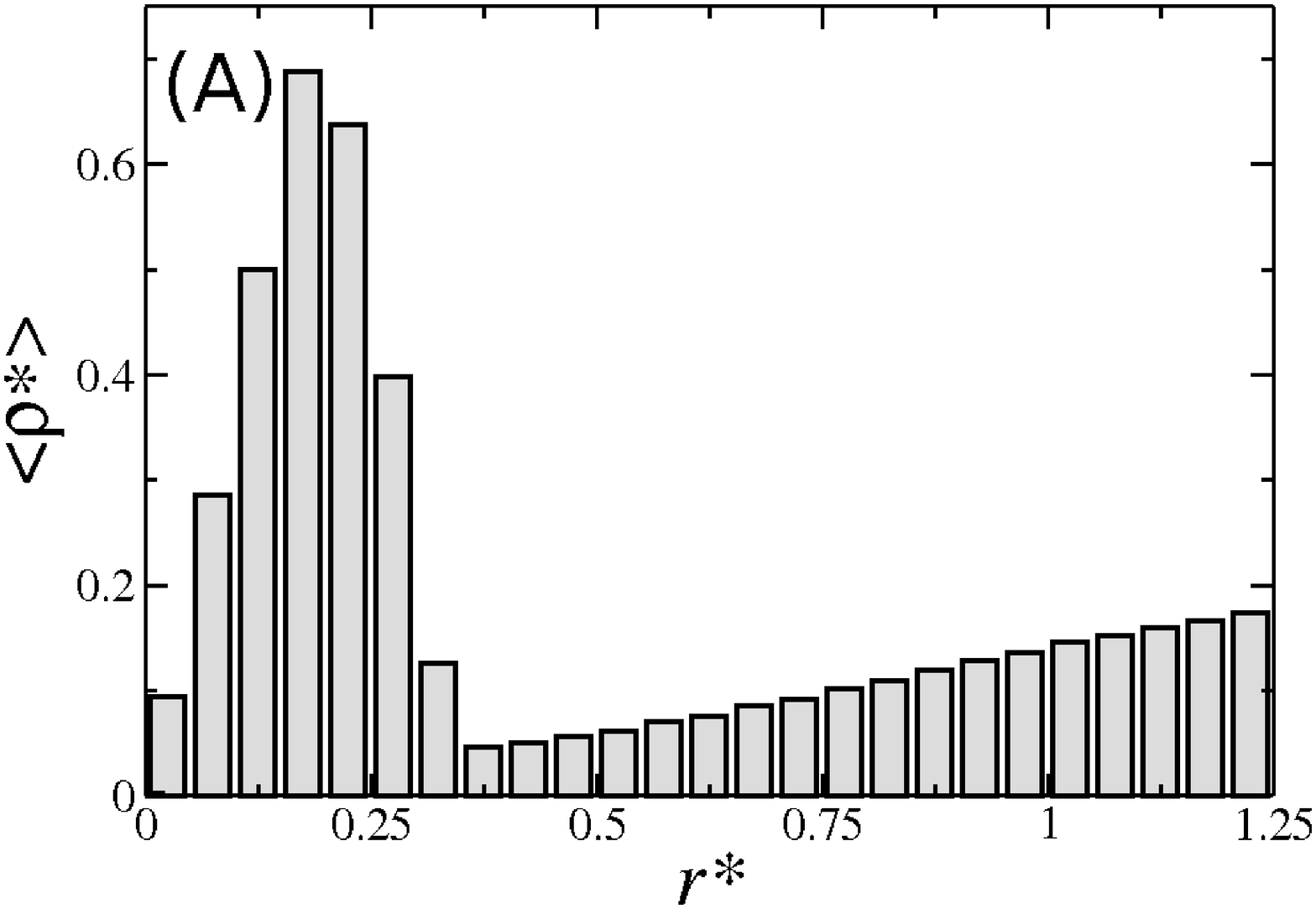}
\includegraphics[width=5cm]{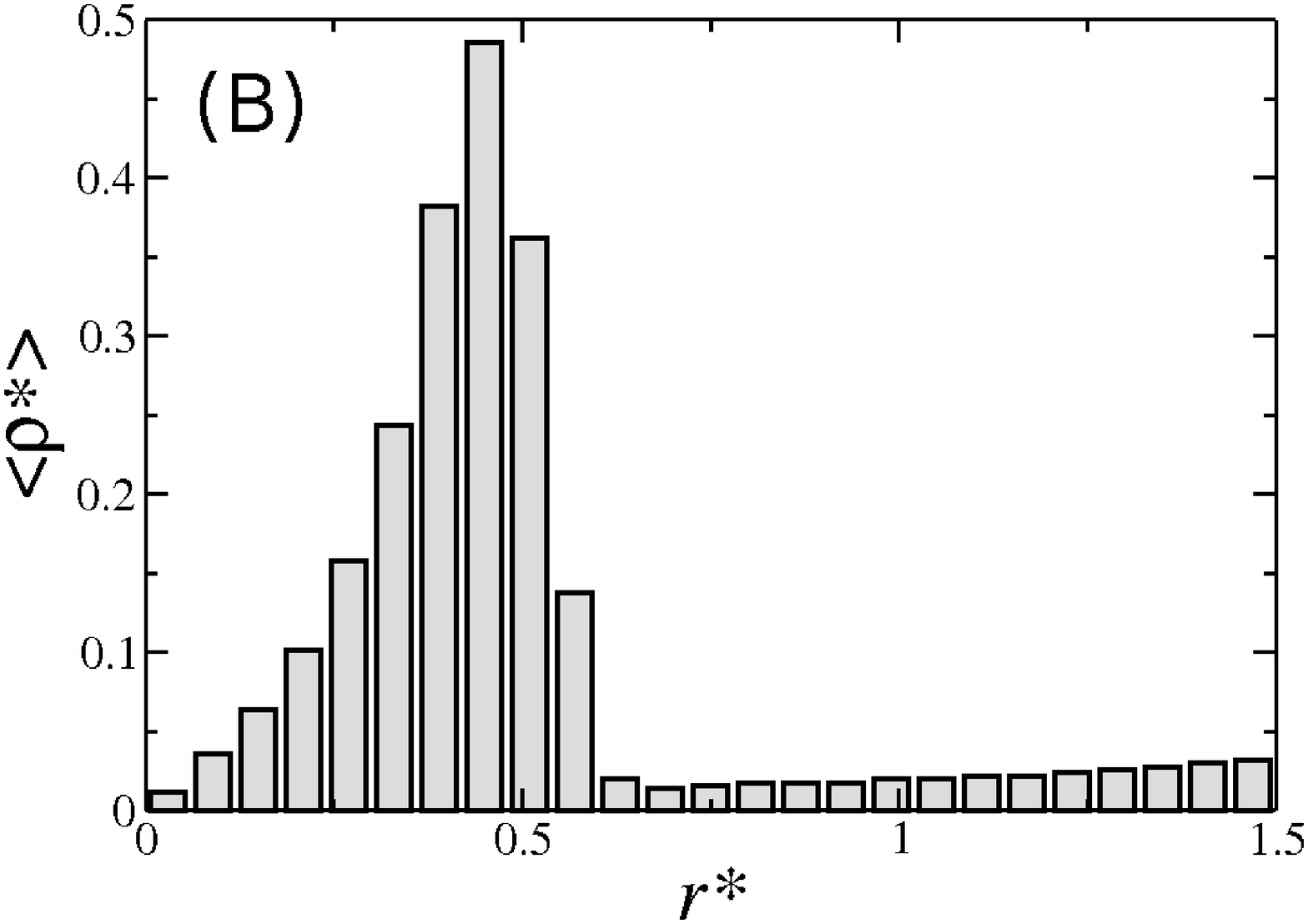}
\includegraphics[width=5cm]{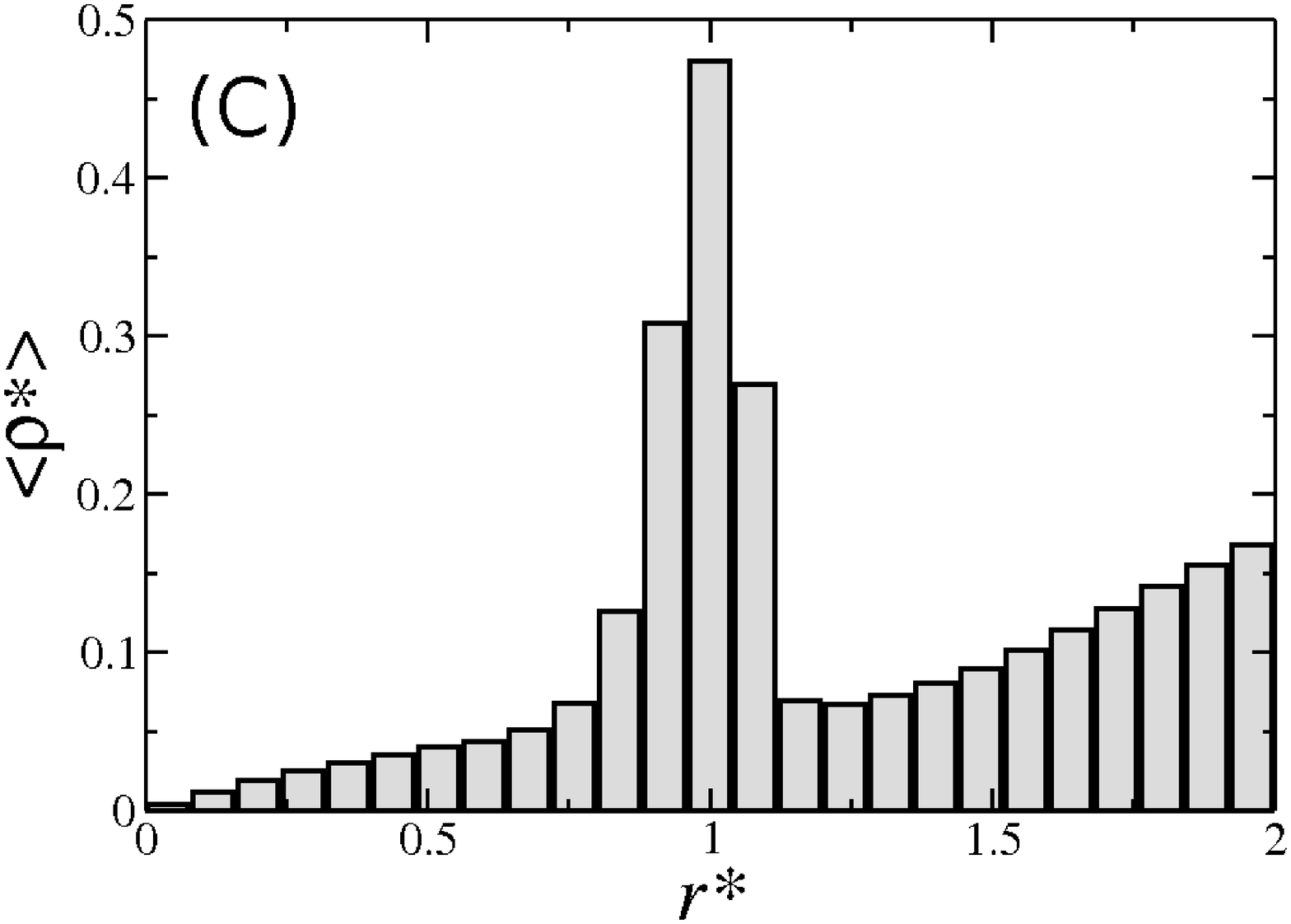}
\includegraphics[width=5cm]{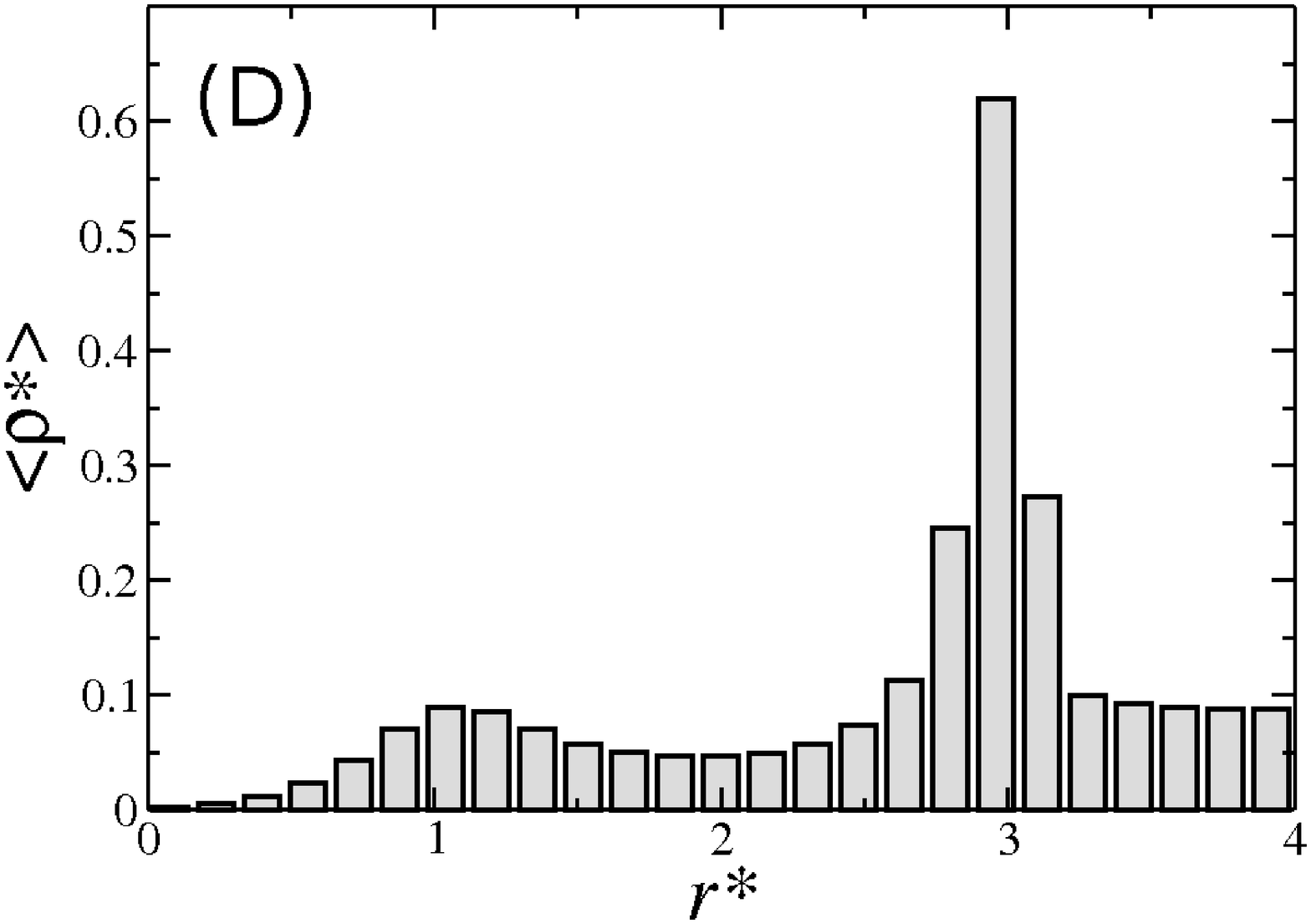}
\includegraphics[width=5cm]{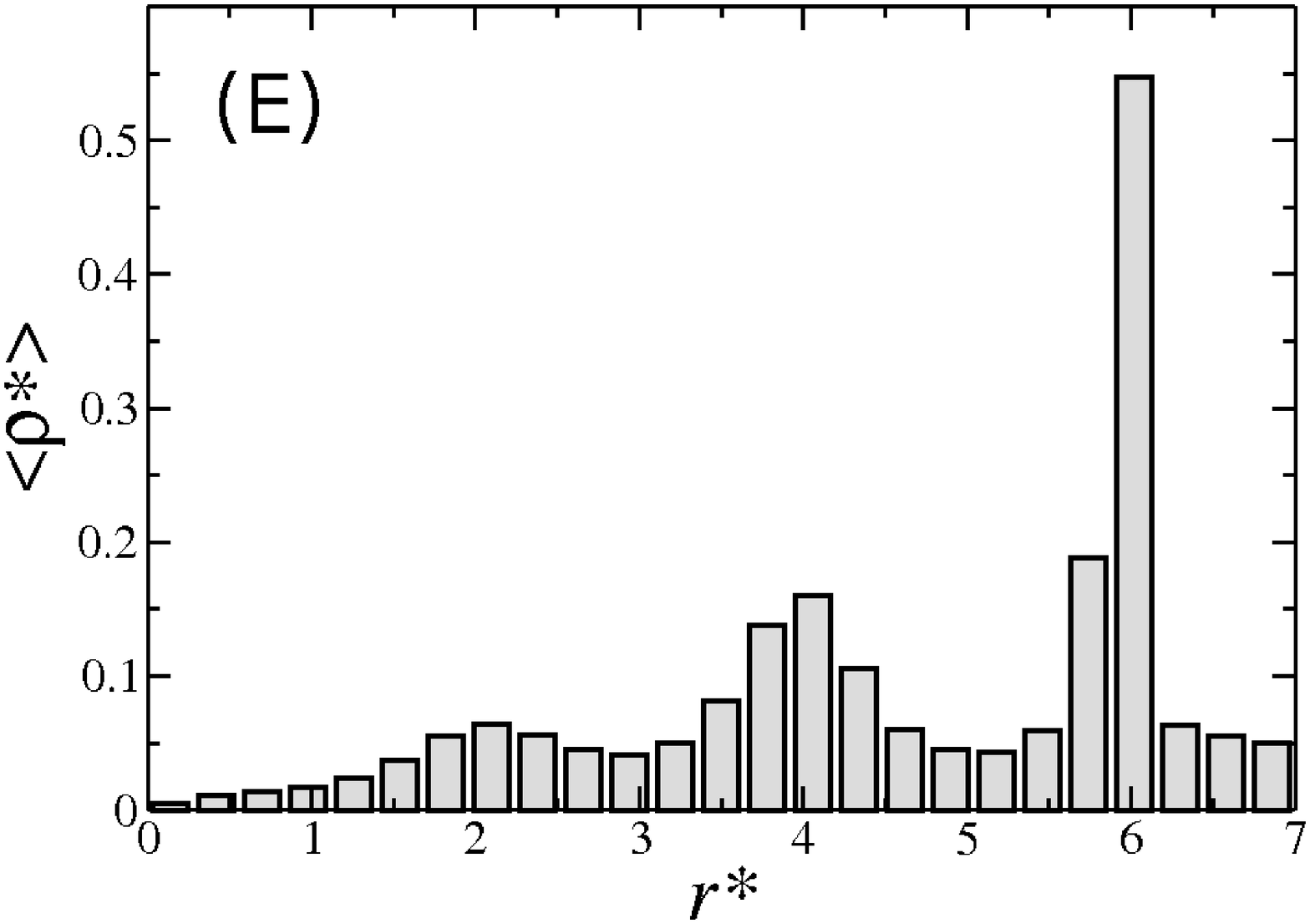}
\includegraphics[width=5cm]{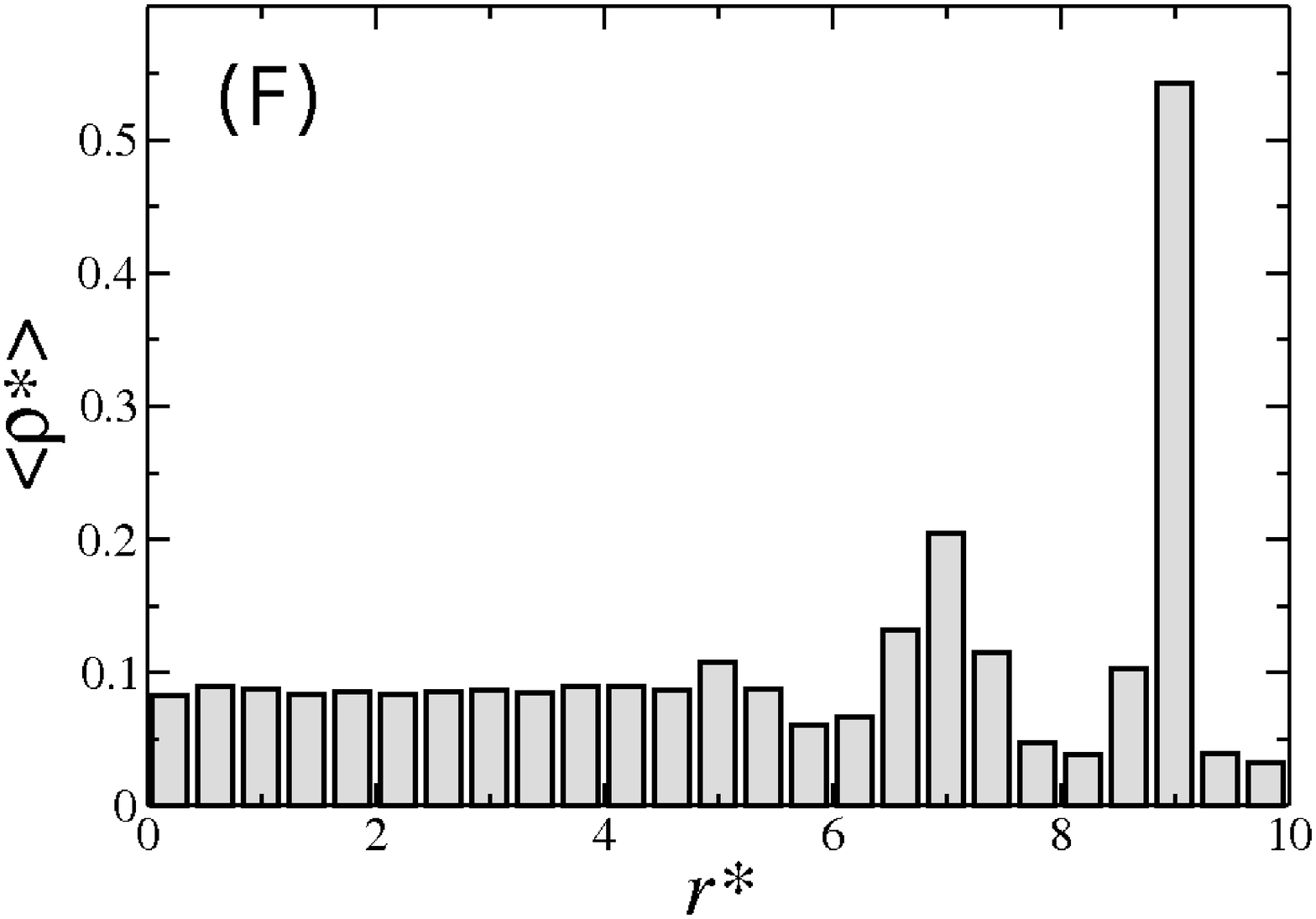}
\end{center}
\caption{Radial density profile for different values of radius: (A) $a^* = 1.25$, (B) $a^* = 1.5$
(C) $a^* = 2.0$, (D) $a^* = 4.0$, (E) $a^* = 7.0$ and (F) $a^* = 10.0$.}
\label{fig:rho.vs.r}
\end{figure}

The behavior of the diffusion coefficient in our model can
be understood by  two complementary ways.
First, by examining the density profile inside the nanotube. 
The density distribution is computed in cylindrical coordinates, 
$r^2 = y^2 + z^2$, where $r=0$ is the center
of the channel. Fig.~\ref{fig:rho.vs.r} illustrates 
the radial density profile versus $r^*$ for the 
channel radii   $a^* = 1.25$, 1.5, 2.0, 4.0, 7.0, and 10.0. 
In all analyzed cases  layering is observed.
Axial layers are also 
observed in simulations for the SPC/E and  TIP4P-EW models
for water confined in nanotubes~\cite{MicroNano05, Mashl03, Mukherjee07, Nanok09, Qin11}. 
In the last  cases,  the presence of layering is 
attributed to the hydrogen bonds and surface effects.
In our model the presence of layering comes as a  result 
of the competition between particle-particle  
and particle-wall interactions. The potential illustrated 
in Fig.~\ref{fig:potential} favors particles  to be at 
least at $r_{pp}^*=2.0$ apart, while
the hydrophobic walls  push
particles away to a distance of at least  $r_{pw}^*= 2^{1/6}$. 
 Consequently, for  $a^* = 1.25$, 1.5, 2.0, 4.0, and 7.0 a number of 
layers equal to 1, 1, 2, 4 and 6  are formed. For $a^*\geq 10$  a continuous distribution 
 emerge. 
For $a^*>2.0$ the system form layers arranged
in distances that minimize the potential energy. 
This indicates that for larger diameters the 
enthalpic contribution for the free energy dominates
over the entropic contribution.
For   $a^*<2.0$ there is only one layer of particles. 
Since the particle-wall interaction is purely repulsive, particles
advance over the wall repulsive region moving ``free'' in the 
radial direction and entropy increases and the 
mobility rises. For $a^*<2.0$ the 
entropic contribution dominates over the enthalpic contribution.

\begin{figure}[ht]
\begin{center}
\includegraphics[width=5cm]{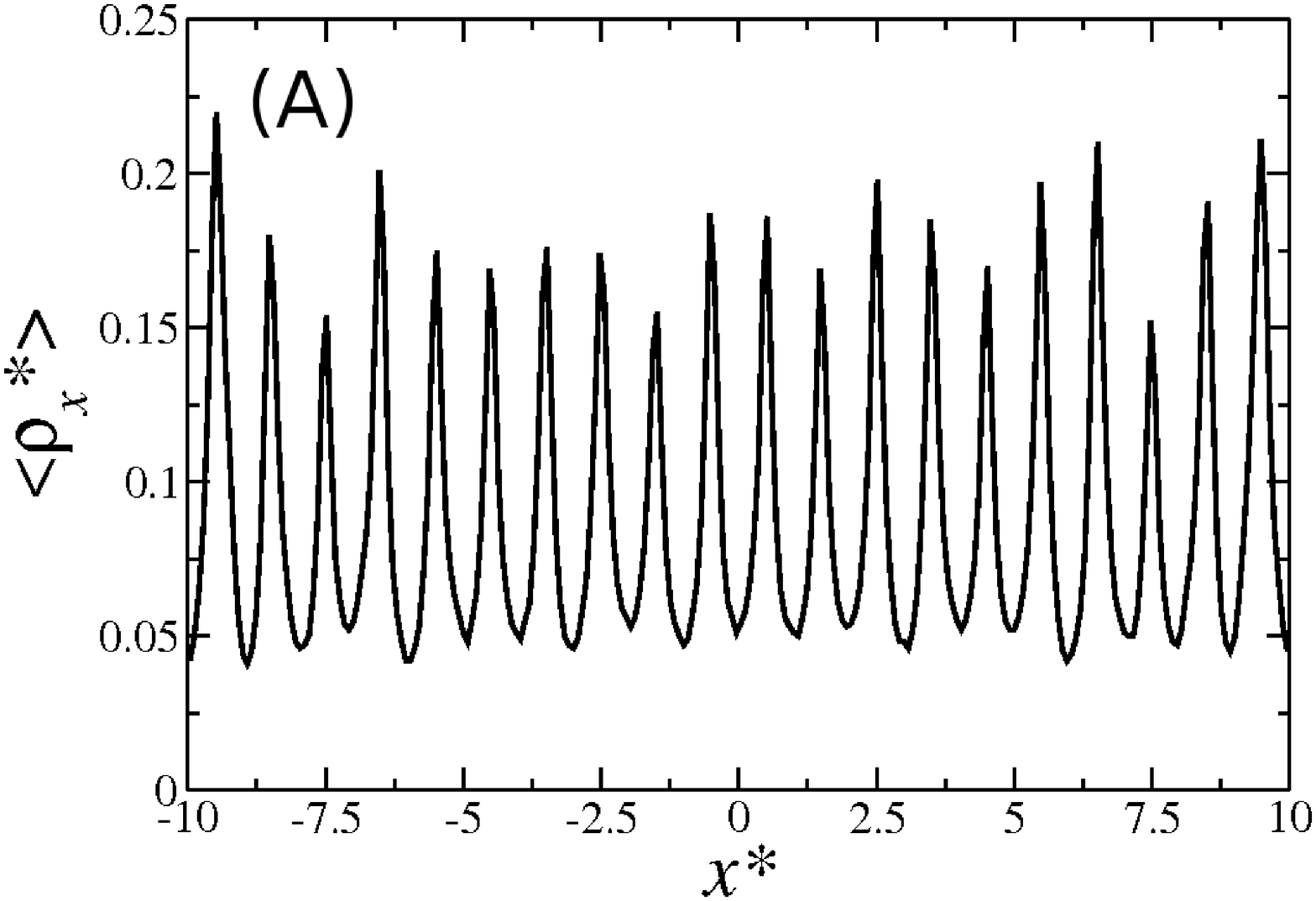}
\includegraphics[width=5cm]{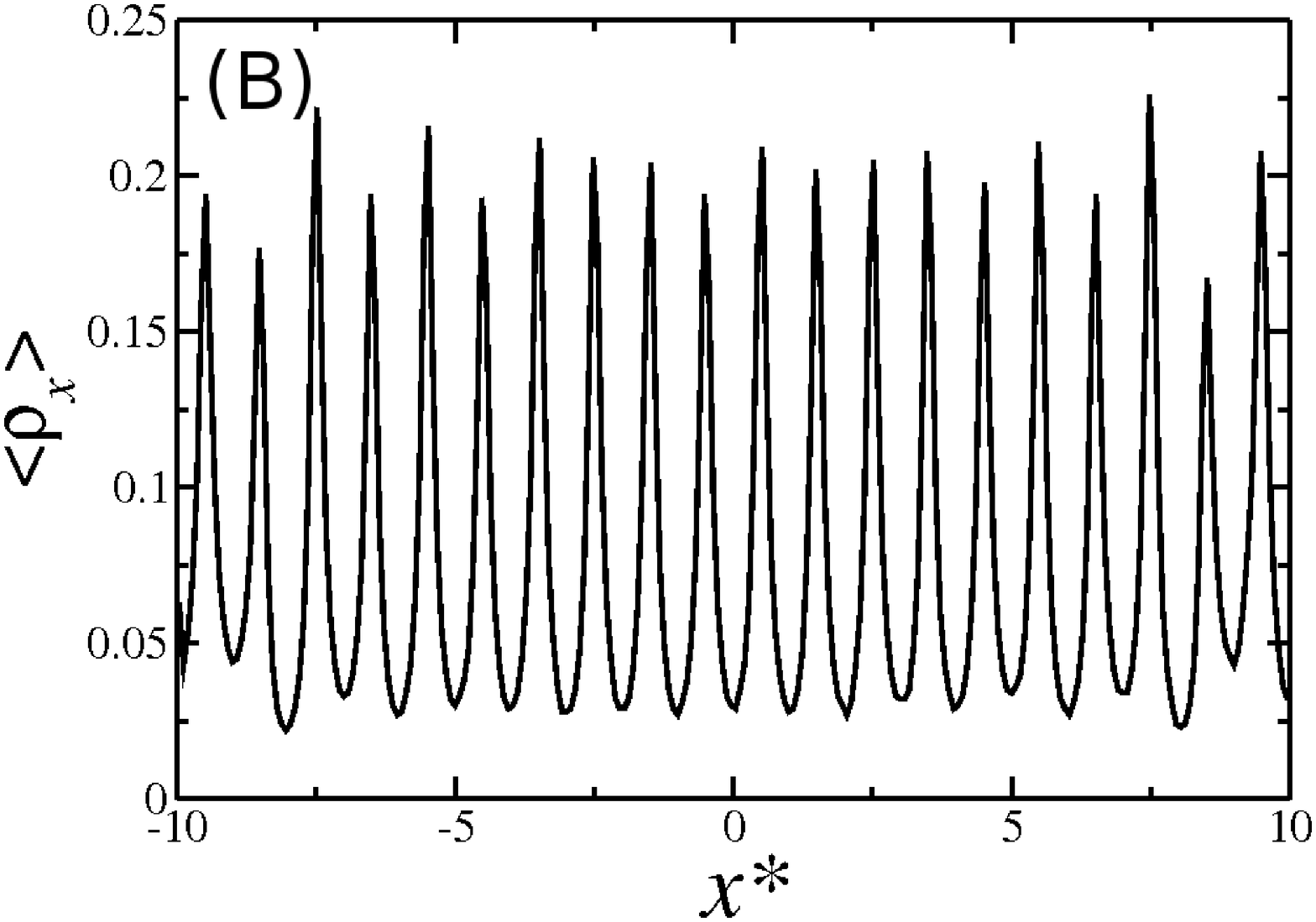}
\includegraphics[width=5cm]{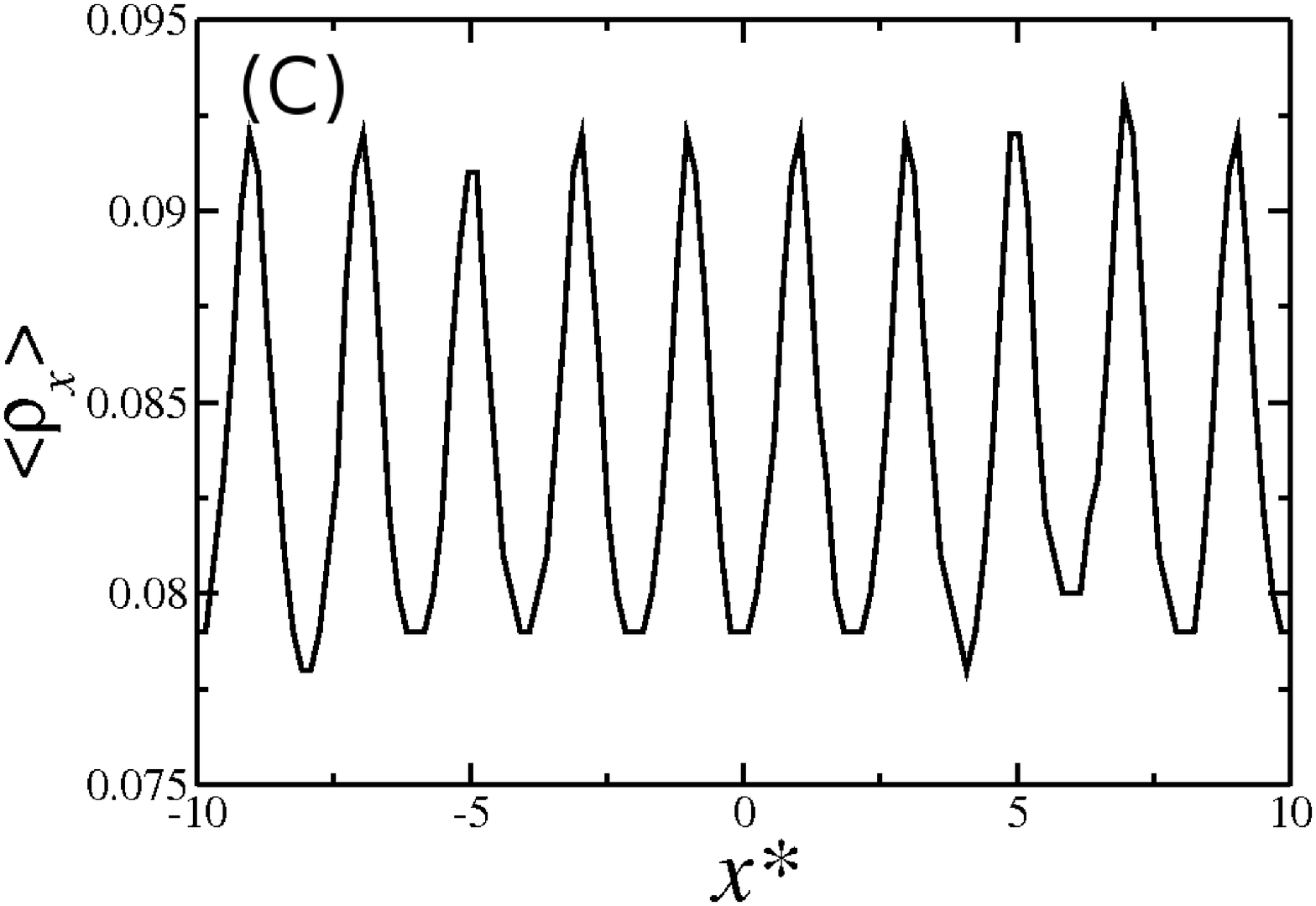}
\end{center}
\caption{Axial density profile for a nanotube with radius  (A)$a^*=1.5$, (B) $a^*=2.0$ and (C) $a^*=7.0$.}
\label{fig:rho.vs.x}
\end{figure}

The decrease in the diffusion 
coefficient as the channel radius is decreased for $a^*>a^*_c=2.0$ is 
associated with the layers formation and particularly with
the correlation between particles in different layers that 
try to move without changing the layer to layer distance. As the number 
of layers increase for $a^*>a^*_c=2.0$ fluctuations allow particles to 
move faster. At  $a^*=a^*_c=2.0$ the diffusion reaches a minimum and the 
system assumes a crystal-like configuration as illustrated
in Figure \ref{fig:rho.vs.x}. Therefore, the 
confinement leads the fluid to a solid-like state, even at values 
of temperature and pressure 
far from solid state 
phase~\cite{MicroNano05, Mashl03, Alexiadis08, Koles06, Nanok09}.

\begin{figure}[nt]
\begin{center}
\includegraphics[width=7cm]{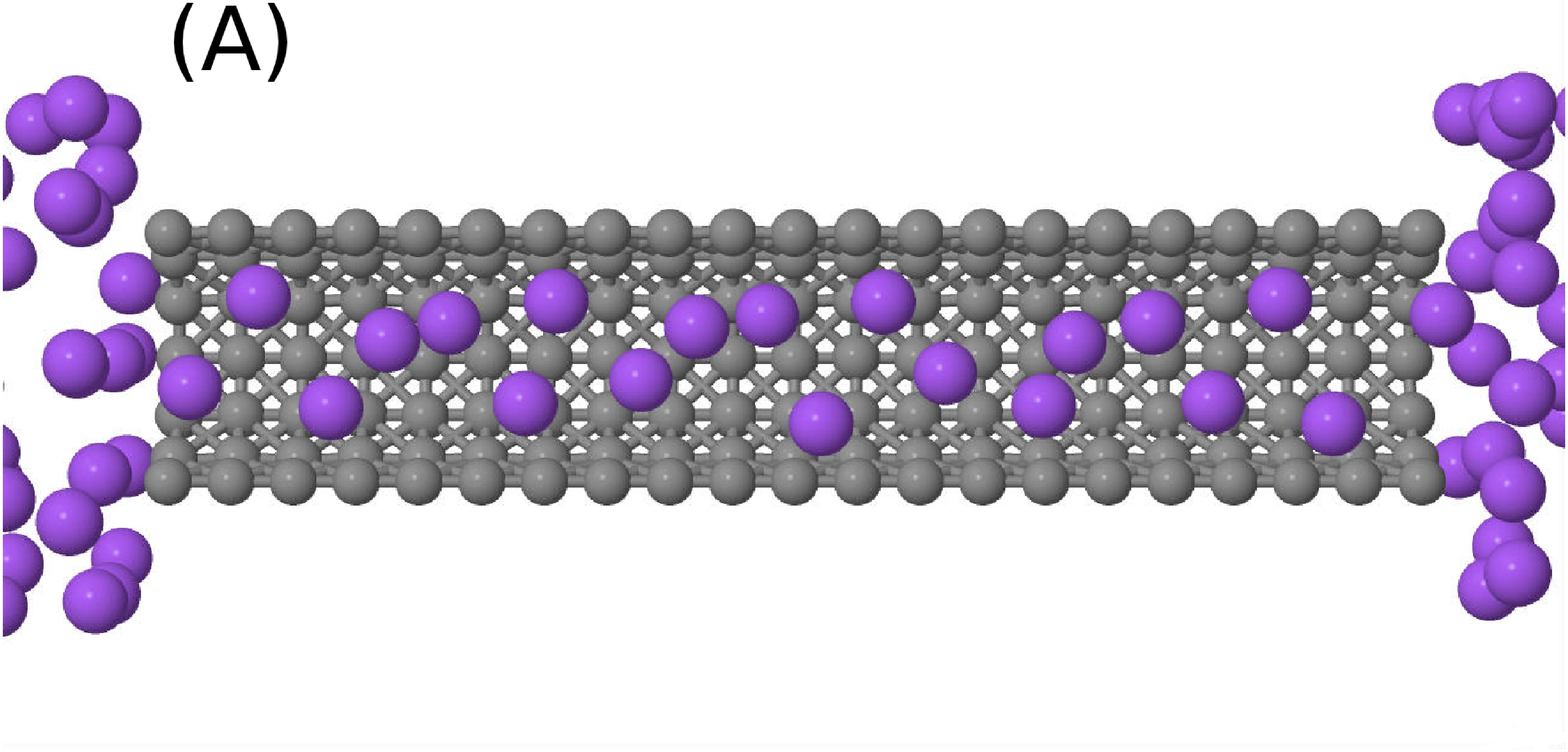}
\includegraphics[width=7cm]{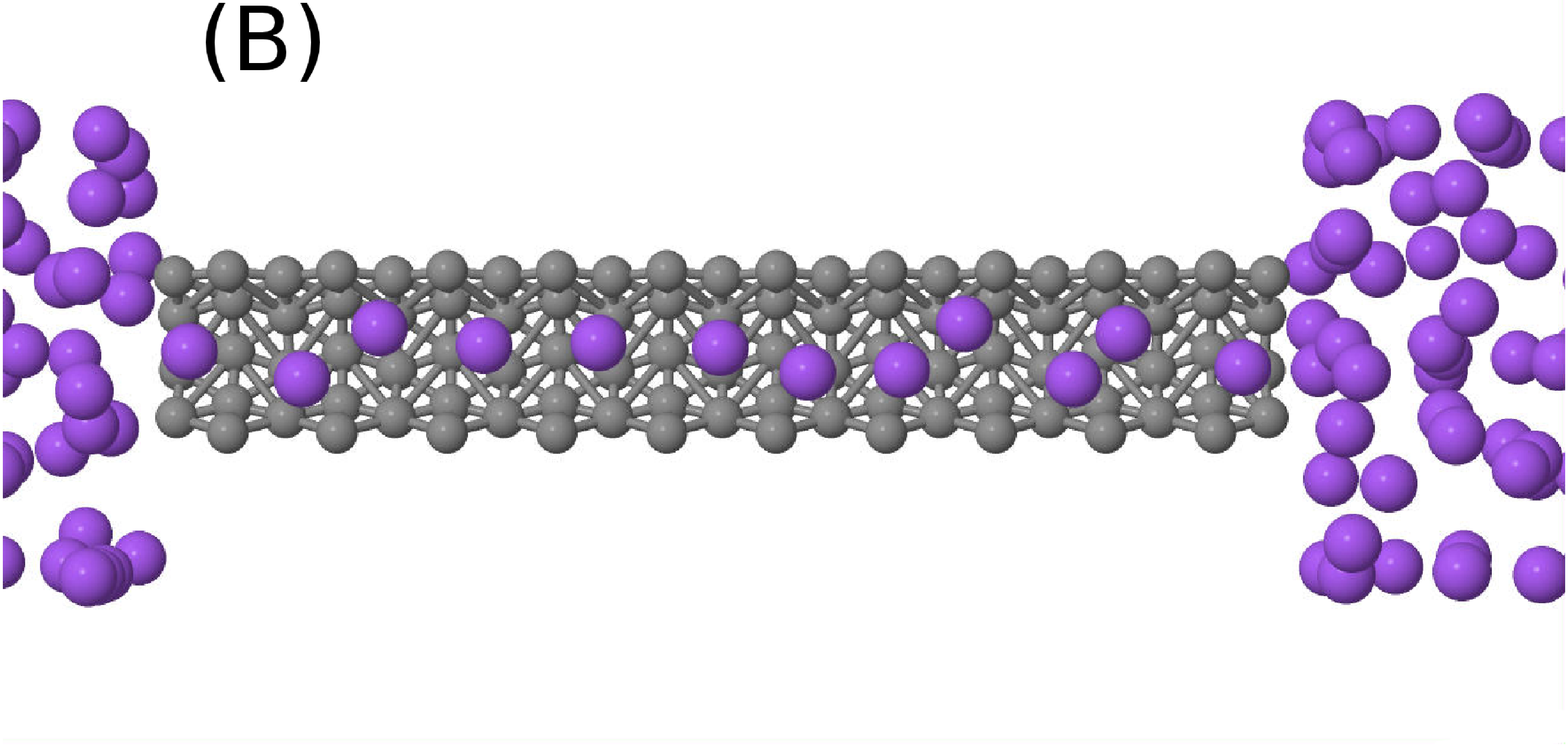}
\end{center}
\caption{Snapshots of the system for (A) $a^*=2.0$ and (B) $a^* = 1.5$.}

\label{fig:snapshots}
\end{figure}

In order to check what happens as the 
channel radius is decreased further, snapshots of the system 
for $a^* = 2.0$ and $a^*= 1.5$ are shown in Fig.~\ref{fig:snapshots}. For 
$a^* = 2.0$  particles form two layers while for $a^* = 1.5$ 
a single layer is observed. The correlations that 
immobilizes particles at the two layers structure disappear
as the single layer is formed and particles can 
diffuse faster for $a^*< 2.0$ by moving from the close packing
(at $r^* = 1.0$) to the minimum of the 
fraction of imaginary modes (at $r^* = 2.0$).

\begin{figure}[ht]
\begin{center}
\includegraphics[width=9cm]{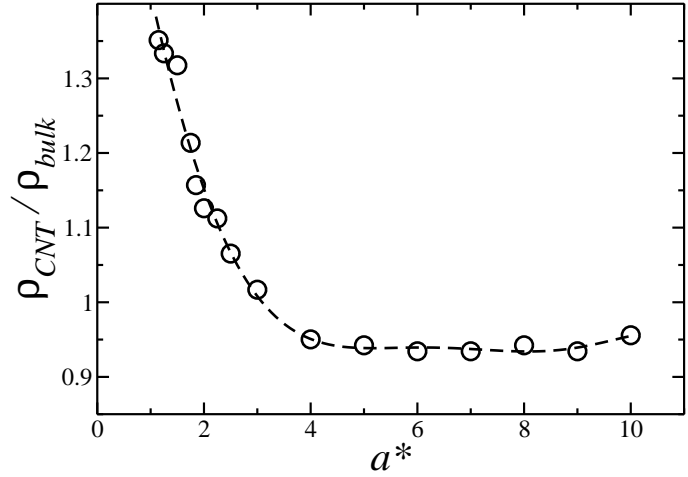}
\end{center}
\caption{Total density inside the nanotube as function of the tube radius for $T^*=0.25$ and $p^*= 0.7$ . The error bars are smaller than the data point. The line is a guide to the eye.}
\label{fig:rho.vs.a}
\end{figure}

Besides the layering in the radial direction, particles
also change their structure in the axial direction.
Fig.~\ref{fig:rho.vs.a} shows the total density inside the nanotube 
as function of the radius.  For $a^* < 4.0$ the density 
of the confined system increases with the decrease of $a^*$. 
This result is qualitatively the same  observed  for recent 
simulations of SPC/E model of water in a nanotube-reservoirs system~\cite{Qin11}.
This can be explained by a change in the axial distance 
between particles. Fig.~\ref{fig:rho.vs.x} for $a^* = 1.5$ 
illustrates that for small radius the preferential axial
distance is the shoulder scale, $x^* = 1.0$,  while in the bulk 
and for larger radius the preferential axial distance is $x^*\approx 2.0$.

\section{Conclusion}
\label{Conclu}

This paper explores the connection between the surface 
interaction, confinement and the presence
of two length scales in the diffusion of
a fluid  in narrow channels. 
The water-like fluid was modeled using a spherically symmetric two 
length  potential, and the confining channel
is modeled as  hard spheres. 
Our system shows  an enhancement
of the diffusion coefficient with the decrease of the 
channel radius for a channel radius
below a certain critical value. This effect arises
from the competition between the confinement, that
accommodate particles at the lower energy length scale, and the 
surface interaction, that pushes  particles
away from the surface generating correlated layers.
For wider  channels the layers 
are accomadated minimizing the 
potential energy forming an organized structure. For
narrow channels the particles form a single layer that
move more freely advancing over the 
wall repulsion and therefore increasing the diffusion.
The mechanism is quite similar to 
the one proposed for water.

In addition we found that below a certain channel radius
the density inside the channel is larger than the 
bulk density. This is explained on basis of
the length scales competition and it is in agreement with
simulations.
Our results indicate that the presence of minimum in 
the diffusion coefficient with the decrease of the 
channel radius is not a property solenely due 
to directional systems such as water but 
can be seen also in spherical symmetric systems
or in system in which directionality would 
not play a relevant role.

\section{Acknowledgments}

This work was partially supported by the CNPq, CAPES, FAPERGS, and INCT-FCx.

\bibliographystyle{aip}

\begin{thebibliography}{10}

\bibitem{Mao00}
Z.~G. Mao and S.~B. Sinnott,
\newblock J. Phys. Chem. B {\bf 104}, 4618 (2000).

\bibitem{Liu10}
Y.-C. Liu, J.~D. Moore, T.~J. Roussel, and K.~E. Gubbins,
\newblock Phys. Chem. Chem. Phys. {\bf 12}, 6632 (2010).

\bibitem{Striolo06}
A.~Striolo,
\newblock Nanoletters {\bf 6}, 633 (2006).

\bibitem{Pikunic03}
J.~Pikunic and K.~E. Gubbins,
\newblock Eur. Phys. J. E {\bf 12}, 35 (2003).

\bibitem{MicroNano05}
G.~E. Karniadakis, A.~Beskok, and N.~R. Aluru,
\newblock {\em Microflows and Nanoflows - Fundamentals and Simulation},
\newblock Springer Science+Business Media Inc, New York, 2005.

\bibitem{watermodels}
M.~Chaplin,
\newblock Water models,
\newblock \url{http://www.lsbu.ac.uk/water/models.html}.

\bibitem{Gallo03}
M.~Rovere and P.~Gallo,
\newblock The European Physical Journal E: Soft Matter and Biological Physics
  {\bf 12}, 77 (2003).

\bibitem{Brovchenko03}
I.~Brovchenko, A.~Geiger, A.~Oleinikova, and D.~Paschek,
\newblock The European Physical Journal E: Soft Matter and Biological Physics
  {\bf 12}, 69 (2003).

\bibitem{Giovambattista09}
N.~Giovambattista, P.~J. Rossky, and P.~G. Debenedetti,
\newblock Phys. Rev. Lett. {\bf 102}, 050603 (2009).

\bibitem{Han10}
S.~Han, M.~Y. Choi, P.~Kumar, and H.~E. Stanley,
\newblock Nature Phys. {\bf 6}, 685 (2010).

\bibitem{B805361H}
T.~G. Lombardo, N.~Giovambattista, and P.~G. Debenedetti,
\newblock Faraday Discuss. {\bf 141},  (2009).

\bibitem{Franzese11}
F.~de~los Santos and G.~Franzese,
\newblock The Journal of Physical Chemistry B {\bf 115}, 14311 (2011).

\bibitem{Gallo12}
P.~Gallo, M.~Rovere, and S.-H. Chen,
\newblock Journal of Physics: Condensed Matter {\bf 24}, 064109 (2012).

\bibitem{Strekalova12}
E.~G. Strekalova, M.~G. Mazza, H.~E. Stanley, and G.~Franzese,
\newblock Journal of Physics: Condensed Matter {\bf 24}, 064111 (2012).

\bibitem{Melillo11}
M.~Melillo, F.~Zhu, M.~A. Snyder, and J.~Mittal,
\newblock The Journal of Physical Chemistry Letters {\bf 2}, 2978 (2011).

\bibitem{Mallamace12}
F.~Mallamace, C.~Corsaro, P.~Baglioni, E.~Fratini, and S.-H. Chen,
\newblock Journal of Physics: Condensed Matter {\bf 24}, 064103 (2012).

\bibitem{Oliveira06a}
A.~B. de~Oliveira, P.~A. Netz, T.~Colla, and M.~C. Barbosa,
\newblock J. Chem. Phys. {\bf 124}, 084505 (2006).

\bibitem{Oliveira06b}
A.~B. de~Oliveira, P.~A. Netz, T.~Colla, and M.~C. Barbosa,
\newblock J. Chem. Phys. {\bf 125}, 124503 (2006).

\bibitem{Silva10}
J.~N. da~Silva, E.~Salcedo, A.~B. de~Oliveira, and M.~C. Barbosa,
\newblock J. Chem. Phys. {\bf 133}, 244506 (2010).

\bibitem{Oliveira10}
A.~B. de~Oliveira, E.~Salcedo, C.~Chakravarty, and M.~C. Barbosa,
\newblock J. Chem. Phys. {\bf 132}, 234509 (2010).

\bibitem{Barraz09}
N.~M.~B. Jr, E.~Salcedo, and M.~C. Barbosa,
\newblock J. Chem. Phys. {\bf 131}, 904509 (2009).

\bibitem{Oliveira08}
A.~B. de~Oliveira, G.~Franzese, P.~Netz, and M.~C. Barbosa,
\newblock J. Chem. Phys. {\bf 128}, 064901 (2008).
vxl
\bibitem{Oliveira08b}
A.~B. de~Oliveira, P.~Netz, and M.~C. Barbosa,
\newblock Eur. Phys. J. B {\bf 64}, 481 (2008).

\bibitem{Oliveira09}
A.~B. de~Oliveira, P.~Netz, and M.~C. Barbosa,
\newblock Europhys. Lett. {\bf 85}, 36001 (2009).

\bibitem{Er06}
J.~R. Errington, T.~M. Truskett, and J.~Mittal,
\newblock J. Chem. Phys. {\bf 125}, 244502 (2006).

\bibitem{Mi06a}
J.~Mittal, J.~R. Errington, and T.~M. Truskett,
\newblock J. Phys. Chem. B {\bf 110}, 18147 (2006).

\bibitem{Kr08}
W.~P. Krekelberg, J.~Mittal, V.~Ganesan, and M.~Truskett, T,
\newblock Phys. Rev. E {\bf 77}, 041201 (2008).

\bibitem{xu:054505}
L.~Xu, S.~V. Buldyrev, N.~Giovambattista, C.~A. Angell, and H.~E. Stanley,
\newblock The Journal of Chemical Physics {\bf 130}, 054505 (2009).

\bibitem{Po92}
P.~H. Poole, F.~Sciortino, U.~Essmann, and H.~E. Stanley,
\newblock Nature (London) {\bf 360}, 324 (1992).

\bibitem{Elimelech11}
M.~Elimelech and W.~A. Philip,
\newblock Science {\bf 333}, 712 (2011).

\bibitem{Hilder11}
T.~A. Hilder, D.~Gordon, and S.~H. Chung,
\newblock Nanomedicine {\bf 7}, 702 (2011).

\bibitem{Liu05}
L.~Liu, S.~H. Chen, A.~Faraone, C.~W. Yen, and C.~Y. Mou,
\newblock Phys. Rev. Lett. {\bf 95}, 117802 (2005).

\bibitem{Mallamace10}
F.~Mallamace, C.~Branca, C.~Corsaro, N.~Leone, J.~Spooren, H.~E. Stanley, and
  S.~H. Chen,
\newblock J. Phys. Chem. B {\bf 114}, 1870 (2010).

\bibitem{Chen06}
S.~H. Chen, F.~Mallamace, C.~Y. Mou, M.~Broccio, C.~Corsaro, A.~Faraone, and
  L.~Liu,Alexiadis
\newblock Proc. Ntl. Acad. Sci. U.S.A. {\bf 103}, 12974 (2006).

\bibitem{Lombardo08}
T.~G. Lombardo, N.~Giovambattista, and P.~G. Debenedetti,
\newblock Faraday Discuss. {\bf 141}, 359 (2008).

\bibitem{Stanley11}
H.~E. Stanley, S.~V. Buldyrev, P.~Kumar, F.~Mallamace, M.~G. Mazza, K.~Stokely,
  L.~Xu, and G.~Franzese,
\newblock J. Non. Crist. Solids {\bf 357}, 629 (2011).

\bibitem{Gelb99}
L.~D. Gelb, K.~E. Gubbins, R.~Radhakrishnan, and M.~S. Bartkowiak,
\newblock Rep. Prog. Phys. {\bf 62}, 1573 (1999).

\bibitem{Mashl03}
R.~J. Mash, S.~Joseph, and N.~R. Aluru,
\newblock Nanoletters {\bf 3}, 589 (2003).

\bibitem{Gordillo00}
M.~C. Gordillo and J.~Mart\'i,
\newblock J. Chem. Phys. Lett. {\bf 329}, 341 (2000).

\bibitem{Kyakuno11}
H.~Kyakuno, K.~Matsuda, H.~Yahiro, Y.~Inammi, T.~Fukukoa, Y.~Miyata, K.~Yanagi,
  H.~Kataura, T.~Saito, M.~Tumura, and S.~Iijima,
\newblock J. Chem. Phys. {\bf 134}, 244501 (2011).

\bibitem{Alexiadis08}
A.~Alexiadis and S.~Kassinos,
\newblock Chem. Rev. {\bf 108}, 5014 (2008).

\bibitem{Holt06}
J.~K. Holt, H.~G. Park, Y.~M. Wang, M.~Stadermann, A.~B. Artyukhin, C.~P.
  Grigoropulos, A.~Noy, and O.~Bakajin,
\newblock Science {\bf 312}, 1034 (2006).

\bibitem{Majumder05}
M.~Majumder, N.~Chopra, R.~Andrews, and B.~J. Hinds,
\newblock Nature {\bf 438}, 44 (2005).

\bibitem{Thomas08}
J.~A. Thomas and A.~J.~H. McGaughey,
\newblock Nanoletters {\bf 8}, 2788 (2008).

\bibitem{Thomas09}
J.~A. Thomas and A.~J.~H. McGaughey,
\newblock Phys. Rev. Lett. {\bf 102}, 4502 (2009).


\bibitem{Qin11}
X.~Qin, Q.~Yuan, Y.~Zhao, S.~Xie, and Z.~Liu,
\newblock Nanoletters {\bf 11}, 2173 (2011).


\bibitem{Ye11}
H.~Ye, H.~Zhang, Y.~Zheng, Z. Zhang,
\newblock Microfluid. Nanofluid. {\bf 11}, 1359 (2011).


\bibitem{Farimani11}
A.~B. Farimani and N.~R. Aluru,
\newblock J. Phys. Chem. B {\bf 115}, 12145 (2011).


\bibitem{Zheng12}
Y.~Zheng, H.~Ye, Z.~Zhang, and H.~Zhang,
\newblock Phys. Chem. Chem. Phys. {\bf 14}, 964 (2012).

\bibitem{Kolesnikov04}
A.~I. Kolesnikov, J.~M. Zanotti, C.~K. Loong, and P.~Thiygarajan,
\newblock Phys. Rev. Lett. {\bf 93}, 035503 (2004).

\bibitem{Koles06}
A.~I. Kolesnikov, C.~K. Long, N.~R. de~Souza, C.~J. Burnham, and A.~P.
  Moravsky,
\newblock Physica B: Condens. Matter {\bf 385}, 272 (2006).

\bibitem{Wang04}
J.~Wang, Y.~Zhu, J.~Zhou, and X.~H. Lu,
\newblock Phys. Chem. ChemJ. Chem. Phys. {\bf 72}, 2384 (1980). Phys. {\bf 6}, 829 (2004).


\bibitem{Alexiadis08_2}
A.~Alexiadis and S.~Kassinos,
\newblock Chem. Eng. Sci. {\bf 63}, 2093 (2008).






\bibitem{Liu08}
Y.~C. Liu, J.~W. Shen, K.~E. Gubbins, J.~D. Moore, T.~Wu, and Q.~Wang,
\newblock Phys. Rev. B {\bf 77}, 125438 (2008).

\bibitem{Nanok09}
T.~Nanok, N.~Artrith, P.~Pantu, P.~A. Bopp, and J.~Limtrakul,
\newblock J. Phys. Chem. A {\bf 113}, 2103 (2009).


\bibitem{Liu05_b}
Y.~Liu, Q.~Wang, and L.~Zhang,
\newblock J. Chem. Phys. {\bf 123}, 234701 (2005).

\bibitem{Su11}
J.~Su and H.~Guo,
\newblock J. Chem. Phys. {\bf 134}, 244513 (2011).


\bibitem{Mukherjee07}
B.~Mukherjee, P.~K. Maiti, C.~Dasgupta, and A.~K. Sood,
\newblock J. Chem. Phys. {\bf 126}, 124704 (2007).

\bibitem{Falk10}
K.~Falk, F.~Sedlmeier, L.~Joly, R.~R. Netz, and L.~Bocquet,
\newblock Nanoletters {\bf 10}, 4067 (2010).

\bibitem{AllenTild}
P.~Allen and D.~J. Tildesley,
\newblock {\em Computer Simulation of Liquids},
\newblock Oxford University Press, Oxford, 1987.

\bibitem{Kell67}
G.~S. Kell,
\newblock J. Chem. Eng. Data {\bf 12}, 66 (1967).

\bibitem{Angell76}
C.~A. Angell, E.~D. Finch, and P.~Bach,
\newblock J. Chem. Phys. {\bf 65}, 3063 (1976).


\bibitem{Andersen80}
H.~C. Andersen,
\newblock J. Chem. Phys. {\bf 72}, 2384 (1980).

\bibitem{LupSmol90}
M.~Lupowski and F.~van Smol,
\newblock J. Chem. Phys. {\bf 93}, 737 (1990).

\bibitem{Thompson99}
A.~P. Thompson and G.~S. Heffelfinger,
\newblock J. Chem. Phys. {\bf 110}, 10693 (1999).


\bibitem{Sa12}
F. de los Santos and G. Frazese
\newblock Phys. Rev. E {\bf 85}, 010602 (2012).

\bibitem{Ku11}
H. Kumar, B. Mukherjee, S-T Lin and C. Dasgupta
\newblock J. Chem. Phys. {\bf 134},  124105(2011).


\bibitem{Pa11}
T. A. Pascal, W. A. Goddard and Y, Jung
\newblock Proc. Nat. Acad. Sci. {\bf 108}, 11794 (2011).



\end{thebibliography}

\end{document}